\documentclass[%
 aip,
amsmath,amssymb,
 reprint,%
]{revtex4-1}

\usepackage{graphicx}
\usepackage{dcolumn}
\usepackage{bm}

\usepackage[utf8]{inputenc}
\usepackage[T1]{fontenc}
\usepackage{mathptmx}
\usepackage{etoolbox}
\usepackage{siunitx}

\makeatletter
\def\@email#1#2{%
 \endgroup
 \patchcmd{\titleblock@produce}
  {\frontmatter@RRAPformat}
  {\frontmatter@RRAPformat{\produce@RRAP{*#1\href{mailto:#2}{#2}}}\frontmatter@RRAPformat}
  {}{}
}%
\makeatother
\begin{document}

\preprint{AIP/123-QED}

\title{Nuclear Recoil Migdal Effect in Liquid Xenon Dark Matter Experiments} 



\author{Jingke Xu}
\email[Corresponding Author: ]{xu12@llnl.gov}
\affiliation{Lawrence Livermore National Laboratory (LLNL), Livermore, CA 94550-9698, USA}
\author{Jeonghwa Kim} 
\affiliation{University of California, Santa Barbara, Department of Physics, Santa Barbara, CA 93106-9530, USA}  
\author{Duncan Adams} \affiliation{C.N. Yang Institute for Theoretical Physics, Stony Brook University, Stony Brook, NY 11794, USA}  
\author{Brian G. Lenardo} \affiliation{SLAC National Accelerator Laboratory, Menlo Park, CA 94025, USA}  
\author{Walter Hugh Lippincott}
\affiliation{University of California, Santa Barbara, Department of Physics, Santa Barbara, CA 93106-9530, USA}
\author{Rouven Essig} \affiliation{C.N. Yang Institute for Theoretical Physics, Stony Brook University, Stony Brook, NY 11794, USA}

\date{\today}

\begin{abstract}
The Migdal effect predicts that a nuclear recoil can be accompanied by detectable atomic ionization or excitation signals, even at the low energies expected from interactions of sub-GeV dark matter particles with atomic nuclei. Liquid xenon-based dark matter experiments have projected substantial sensitivity gains to light dark matter based on this effect, underscoring the importance of its direct characterization in xenon. In this Letter, we draw on our theoretical and experimental  studies of nuclear recoil Migdal interactions to discuss their predicted characteristics and corresponding observable signatures in liquid xenon detectors. We examine the challenges of directly observing Migdal signals using neutron-induced xenon recoils and outline possible measurement strategies and necessary background mitigation measures to allow a definitive confirmation of the Migdal effect in liquid xenon.
\end{abstract}


\maketitle 


Liquid xenon-based dark matter experiments have demonstrated unmatched sensitivities to dark matter candidates with masses above a few GeV/c$^2$ through their extremely low intrinsic radioactivity levels, strong self-shielding against external backgrounds, and mass scalability. More importantly, dual-phase xenon time projection chambers (TPCs) can collect both scintillation (S1) and ionization (S2) signals produced by particle interactions in the liquid. The combination of these two signal channels provides enhanced energy resolution, 3D position sensitivity, and powerful discrimination between nuclear recoil (NR) and electron recoil (ER) interactions down to the keV level. 
Xenon TPC experiments including LZ~\cite{LZ2024_WS}, XENONnT~\cite{XenonNT2025_NR} and PandaX~\cite{PandaX2025_NRDM} have dominated the direct searches for medium-to-high-mass dark matter particles, with the next-generation experiment XLZD designed to fully explore the parameter space for weakly interacting massive particles (WIMP) above the neutrino fog~\cite{XLZD2024_DesignBook}.

Two recent developments have enabled xenon TPCs to also gain competitive sensitivities to sub-GeV dark matter candidates. The first is the analysis strategy of only using ionization signals (``S2-only'') to increase an experiment's acceptance of low-energy events at the cost of elevated background levels~\cite{Essig2012_SubGeVXENON10,Essig2012_SubGeVDM,LUX_Migdal_2021,pandax4t_s2o}.
Traditional dark matter searches in xenon TPCs are limited to energy thresholds of $\gtrsim$3~keV for NR events and $\gtrsim$1~keV for ERs by the low photon detection efficiency ($\sim$10\%) and the multi-photon requirement for valid S1s. S2-only searches, however, leverage the high efficiency to detect ionization electrons (50-100\%), resulting in a possible NR analysis threshold as low as $\sim$200~eV~\cite{Lenardo2019_XeNR}. When considering dark matter interactions with atomic electrons that produce low-energy ERs, signals as low as $\sim$15~eV can be probed~\cite{Essig2012_SubGeVXENON10,XENONnT2025_S2o}. 
The second development is the renewed interest in the Migdal effect~\cite{Migdal_1941,Ibe2017_Migdal,Cox2022_Migdal,Essig2020:Migdal,Liu2020_Migdal,Baxter2020:Migdal}. A.B. Migdal proposed in 1941 that sudden changes in the state of an atomic nucleus following radioactive decays or kinematic recoils can cause the atom to be excited or ionized~\cite{Migdal_1941}. The effect was observed for beta decays in 1954~\cite{Boehm1954_BetaMigdal} and for alpha decays in 1975~\cite{Rapaport1975_AlphaMigdal}. Due to difficulties in definitively detecting low-energy NR signatures, the NR Migdal effect was only recently observed in a gas medium~\cite{Yi2026_Migdal}. In dark matter search experiments, the NR Migdal effect can be viewed as an inelastic scatter between an dark matter particle and an atom~\cite{Ibe2017_Migdal}, which allows a larger fraction of the dark matter energy to be transferred to the atom, boosting the signal strength especially in low-mass dark matter interactions. 

\begin{figure}[t!]
\centering
  \includegraphics[width=\linewidth]{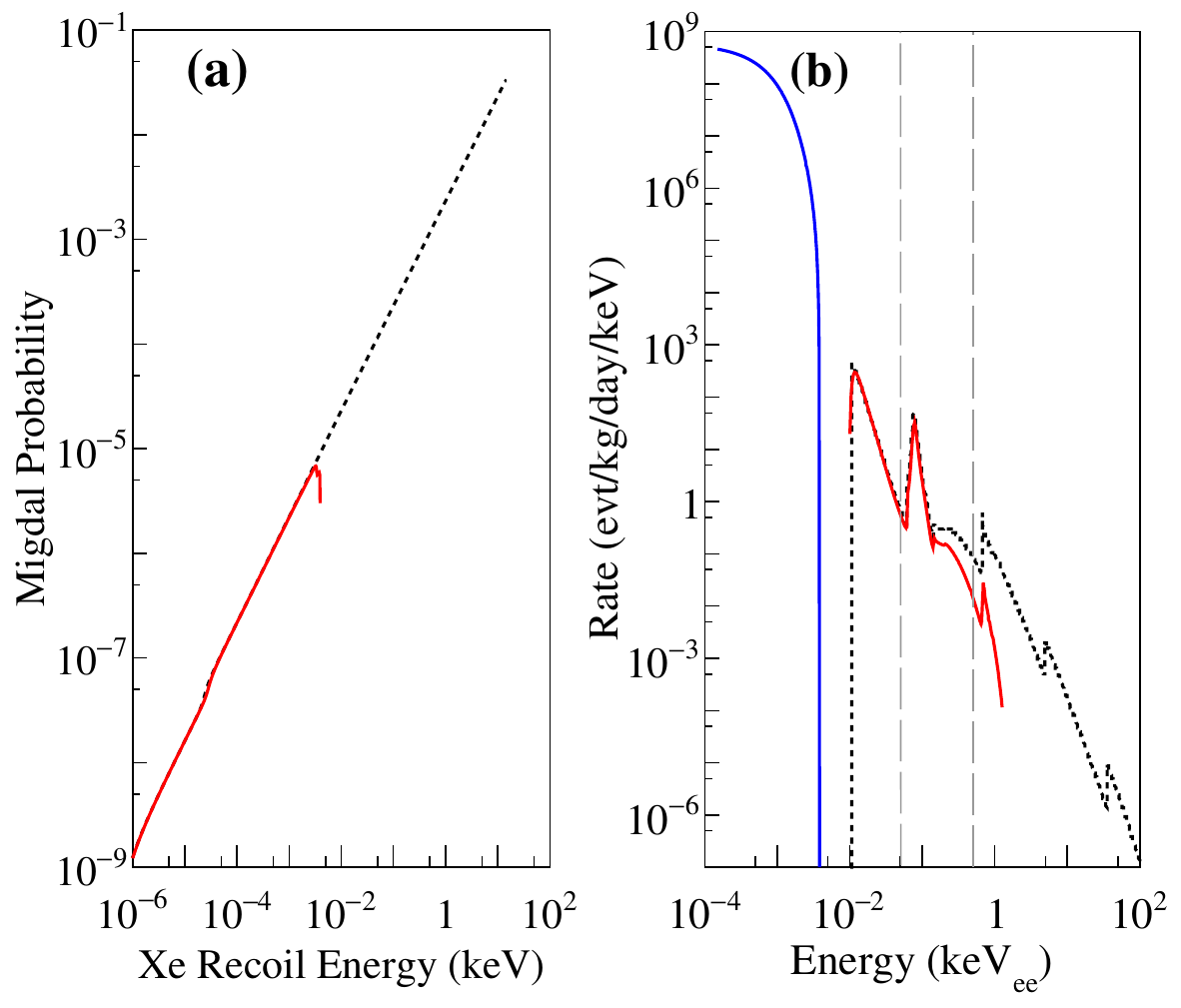}
\caption{(a) Predicted Migdal interaction probability for scatters  of 0.5~GeV (solid red) and 40~GeV (dotted black) dark matter particles with 0.001$c$ initial speed with xenon. (b) The pure NR (blue) and Migdal interaction (red) spectra for 0.5~GeV dark matter interactions with xenon; the dotted black spectrum is the scaled Migdal ER spectrum for interactions without energy constraints in the system; the vertical dashed gray lines are the approximate S2-only (lower energy) and S1+S2 (higher energy) analysis thresholds in typical xenon TPCs. The NR energy and the NR component of the 0.5~GeV Migdal spectra have been scaled down by a factor of 0.15 to reflect the lower signal yield of NRs relative to ERs in liquid xenon. }
\label{fig:rate}
\end{figure}

In the energy region relevant to dark matter searches, the predicted Migdal probability is approximately linear with the NR energy as a result of the linear dependence of the Migdal interaction matrix element on the NR velocity. The black (red) line in Fig.~\ref{fig:rate}a shows the total Migdal interaction probabilities for xenon NRs produced by a 40~GeV (0.5~GeV) dark matter particle of 0.001$c$ initial speed, representative of galactic dark matter velocities. The linear relationship breaks down for 0.5~GeV dark matter scatters when the NR energy approaches its kinematic endpoint at $\sim 0.01~\rm{keV}$, which diminishes energy available for Migdal excitation or ionization. 
In particular, it takes a higher energy for an inner shell electron to be excited or ionized, thus the reduction in Migdal probability is the most prominent for inner shells when the total available kinematic energy is constrained. 
The Migdal energy spectrum for 0.5~GeV dark matter particles scattering with xenon is shown in red in Fig.~\ref{fig:rate}b, where a significant loss of M-shell signals is predicted compared to the spectrum with no kinematic energy constraints (dotted black). 
Similarly, no L- or K-shell Migdal signals are expected for 0.5~GeV dark matter interactions with xenon because the available dark matter particle energy is below the xenon atomic energy for these shells. 
In the absence of kinematic energy constraints, the fraction of K-, L-, M-, N- and O-shell Migdal interactions is approximately 5$\times10^{-5}$, 2$\times10^{-3}$, 0.04, 0.31, and 0.64, respectively~\cite{Ibe2017_Migdal}. 

As illustrated in Fig.~\ref{fig:rate}b, S2-only and S1+S2 analyses in xenon TPCs have typical thresholds of $\leq$0.1~keV electron equivalent energy (keV$_{ee}$) and $\sim$1~keV$_{ee}$, respectively. This allows N-shell (via S2-only), M-, L- and K-shell Migdal interactions to be detected with relatively high efficiencies even when the concurrent NR signals fall below the TPC threshold. Meanwhile, the extremely low background levels and large exposures of xenon TPC experiments partially compensate for the low interaction rates of dark matter through the Migdal channel, allowing them to be observed with statistical significance. As such, the Migdal effect has the potential to greatly enhance the light dark matter sensitivity of existing xenon TPCs, especially in combination with low energy-threshold S2-only analyses~\cite{LUX2019:Migdal,LUX_Migdal_2021,XENON1T2019:Migdal,Essig2020:Migdal,PandaX2023_Migdal}. 
These substantial sensitivity gains remain to be validated with a direct NR Migdal effect confirmation in liquid xenon.

Several attempts to experimentally measure the NR Migdal effect have recently been made using neutron scatters~\cite{UKMIGDAL2022,Xu2024_Migdal,LZ_Bang_phd,LZ_Vaitkus_phd,Nakamura2020_Migdal,Yi2026_Migdal}. Neutrons---like dark matter particles---are free of electric charge and primarily interact with atomic nuclei. Therefore, they can be used in a laboratory to produce NR events with the desired energy and directional distributions. This technique is widely used in the calibration of dark matter detectors~\cite{Xu2023_NRCalibration,Lenardo2019_XeNR,LUX2025_NR,CDMS2023_SiNR}, and also creates the conditions for the Migdal effect to be tested. For example, $\mathcal{O}$(10)~keV NRs produced by scatters of 2.5~MeV neutrons were used in the first reported NR Migdal detection where Migdal signals were observed as co-located NR and ER tracks in a gas detector medium~\cite{Yi2026_Migdal}. 

\begin{figure}[t!]
\centering
  \includegraphics[width=\linewidth]{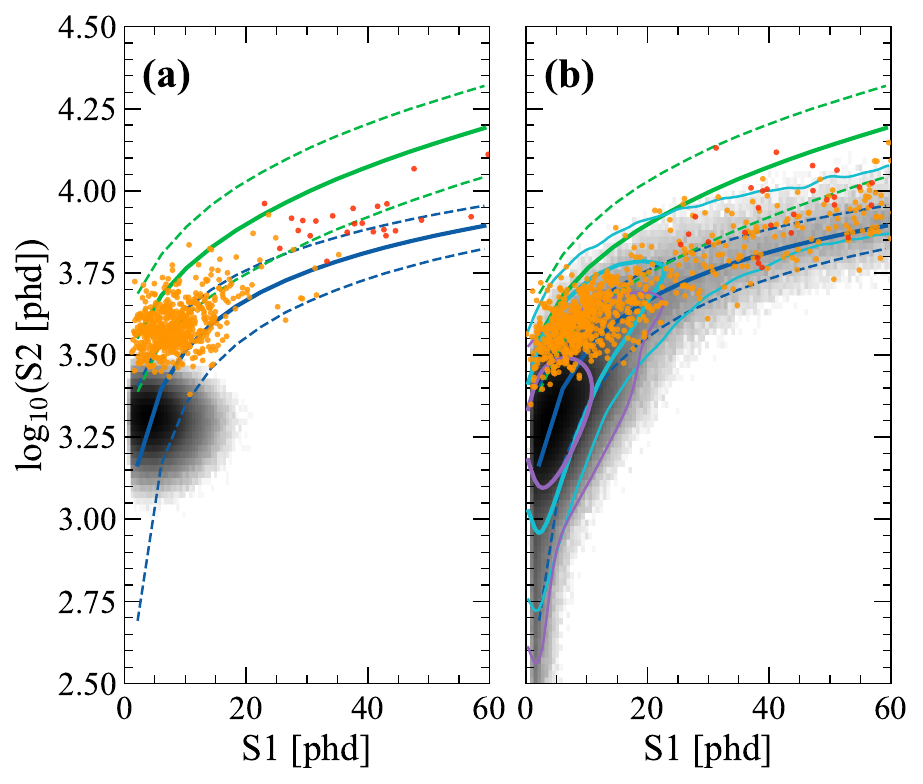}
\caption{(a) Simulated 7~keV xenon NRs (gray) and predicted M-shell (orange) and L-shell (red) Migdal signals in the ideal case; (b) similar to the left but for simulated 7.0$\pm$1.5~keV xenon recoils and neutron multiscatter backgrounds produced by scattering 14.1~MeV neutrons off a small xenon TPC at $\sim$15$^\circ$ angle~\cite{Xu2024_Migdal}; TPCMS (purple contours, 1$\sigma$ and 2$\sigma$) and PMS backgrounds (cyan contours, 1$\sigma$ and 2$\sigma$) are observed in the Migdal signal region. The NR and ER yield values are calculated using NEST~\cite{Szydagis2025_NEST} v2.4.0 at an electric field of 200~V/cm, and their 10-50-90\% bands are shown in blue and green, respectively. The TPC is assumed to have an S1 collection efficiency of 0.1~ detected photon (phd)/photon and an S2 gain of 45~phd/electron. Both simulations assume 10$^6$ xenon recoils with S1$\ge$1~phd. }
\label{fig:sigmodel}
\end{figure}

In the popular dark matter search medium of liquid xenon, however, low-energy NR and ER tracks are too short to be resolved, and alternative event topologies are required to identify Migdal signals. 
Fig.~\ref{fig:sigmodel}a shows the simulated S1 and S2 distributions of M-shell (orange dots) and L-shell (red dots) Migdal events for monoenergetic 7~keV xenon recoils without constraining the available energy for Migdal interactions (gray histogram). 
In L- and higher-shell xenon Migdal interactions, the resulting atomic vacancies predominantly decay through emission of Auger electrons. 
Therefore, the high stopping power of liquid xenon causes virtually all energy in such Migdal interactions to be locally deposited. 
This high concentration of ionizing particles suppresses the observable ionization signal strength by enhancing electron-ion recombination because the ions and electrons produced by one primary particle are simultaneously exposed to, and are allowed to combine with, electrons and ions from other ionizing particles. 
Based on measurements by XELDA~\cite{Temples2021_127Xe} and LZ~\cite{LZ2025_EC}, the total S2 yields of M- and L-shell atomic relaxations are suppressed to $\sim$90\% of the corresponding beta values. The combined Migdal signal is also subject to enhanced recombination between the co-located NR and ER tracks, a new effect studied in our recent work~\cite{Xu2025_Recombination}. 

After incorporating these signal suppression effects, our model predicts that Migdal interactions can still be distinguished from pure NRs in the S1-S2 parameter space. 
Specifically, the $\sim$1~keV M-shell Migdal ER energy produces a similar number of ionization electrons to that of a 7~keV NR, 
and also modestly contributes to the photon signal. The cross-track ER-NR recombination effect causes an average of 15 ionization electrons to be converted to photons in the overall signal, which increases the observable S1 size by 1.5 photons detected (phd) with the assumed light collection efficiency of 10\%.  
Ultimately, the NR Migdal events are predicted to produce significantly larger S2 sizes than pure NRs with a modest S1 enhancement, occupying a parameter space between typical ER and NR events. Importantly, this signature is distinct from yield fluctuations of pure NRs, which could increase the observed S1 or S2 strength at the cost of lowering the other. 
As a result, Migdal events in liquid xenon can potentially be identified in an experiment where relevant backgrounds are adequately mitigated.

We carried out a Migdal search experiment with approximately half a million 7.0$\pm$1.5~keV xenon recoils produced by scatters of 14.1~MeV neutrons off a compact xenon TPC at $\sim$15$^\circ$ angle~\cite{Xu2024_Migdal}. In addition to creating semi-monoenergetic NRs, the neutron tagging method also strongly suppresses non-neutron-induced backgrounds through a time coincidence, suppressing ER backgrounds to negligible levels. However, the data identified a dangerous background from neutron multiscatters. 
Fig.~\ref{fig:sigmodel}b illustrates the two types of multiscatter backgrounds using a full simulation of the measurement setup. Multiscatter events with the neutron interacting only once in the TPC but depositing additional energy in passive detector materials, dubbed PMS, produce a continuous NR spectrum (cyan contours), the high energy component of which partially overlaps with Migdal signals. 
Multiscatter events with two or more neutron interactions in the TPC, dubbed TPCMS (purple contours) produce events that populate between the NR and ER parameter spaces, also overlapping with Migdal signals. 
Despite the use of high-energy neutrons, a small TPC volume and stringent cuts to suppress neutron multiscatters, rates of remaining TPCMS and PMS backgrounds in the signal region of interest are comparable to that of the predicted Migdal signals~\cite{Xu2025_Recombination}. 
It should be noted that our initial Migdal signal search did not consider the enhanced cross-track recombination effect and predicted larger S2 sizes for Migdal signals than for neutron multiscatter backgrounds, creating an apparent tension with experimental data~\cite{Xu2024_Migdal}. 

\begin{figure}[t!]
\centering
  \includegraphics[width=\linewidth]{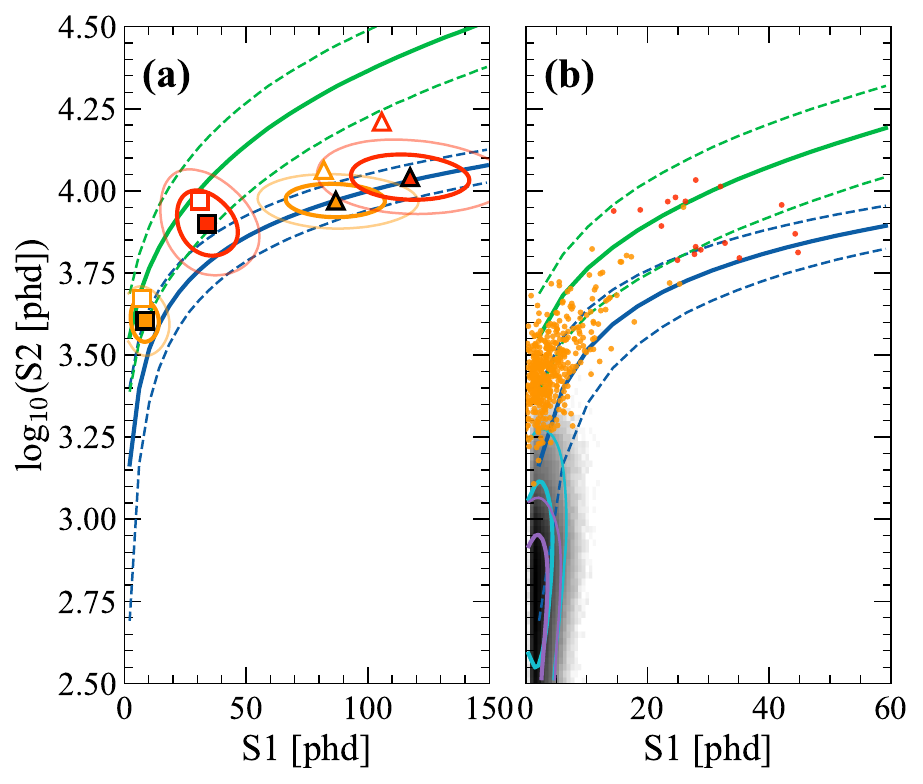}
\caption{(a) Predicted S1-S2 distributions for M-shell (orange) and L-shell (red) Migdal signals accompanying 7~keV (squares) and 60~keV (triangles) xenon NRs before (hollow markers) and after (filled markers) considering cross-track NR-ER recombination enhancement; the 1- and 2$\sigma$ signal contours with enhanced recombination are also shown. (b) Simulated Migdal signal and background distributions for 0--4.5~keV xenon recoils produced by scatters of 100~keV neutrons off the small LLNL xenon TPC; the color coding and band structures are the same as used in Fig.~\ref{fig:sigmodel}. }
\label{fig:hlenergy}
\end{figure}

The predicted Migdal signals illustrated in Fig.~\ref{fig:sigmodel}b contain a population in the large-S1 region that is nearly degenerate with high energy NRs. 
These are primarily Migdal interactions associated with high-energy PMS NR events that have higher Migdal probabilities (Fig~\ref{fig:rate}a). 
This putative benefit of high-energy xenon NRs for Migdal signal searches, however, is largely negated by the cross-track recombination effect. 
Fig.~\ref{fig:hlenergy}a further illustrates the effect of enhanced recombination between the Migdal ER and NR components on the overall signal. For both M-shell and L-shell, higher energy NRs tend to create stronger S2 suppression, bringing the final signal closer to the pure NR population. For NRs above 50~keV, the resulting Migdal signals almost fully coincide with pure NRs, inhibiting their identification in the presence of higher-energy NR backgrounds.  
A Migdal search could focus on the most energetic NRs allowed by the neutron energy to avoid higher energy NRs, but TPCMS and S1 fluctuations may still obscure the signals.  
Additionally, a strong recombination effect for high energy NRs implies a large uncertainty in the signal model, which is difficult to quantify due to lack of experimental data for model validation.

The aforementioned experimental difficulties suggest that low-energy NR events, which are inherently insensitive to the recombination enhancement effect, are preferred for a robust characterization of the xenon NR Migdal effect. 
Such measurements, however, may face two other challenges. 
First, the probability of Migdal interactions decreases significantly for low-energy NRs (Fig.~\ref{fig:rate}a), thereby requiring higher NR statistics and stronger background rejection. 
Second, low-energy NRs are most practically produced by low energy neutrons which have short mean free paths in liquid xenon or other materials and may produce higher rates of multiscatter backgrounds. 
These two challenges can be simultaneously addressed with an experiment using $<$150~keV neutrons.  A neutron of this energy transfers a maximum of 4.5~keV to xenon in a single elastic scatter, eliminating high-energy PMS backgrounds. This low single-scatter energy transfer also prevents TPCMS from producing large ionization responses in the Migdal signal region. 
Such low-energy NR events have a small energy spread by definition, alleviating the need to tag scattered neutrons and thus boosting event statistics. 

Fig.~\ref{fig:hlenergy}b shows the predicted M- and L-shell Migdal signal distributions (orange and red dots) for xenon recoils (gray histogram) produced by 100~keV incident neutrons, where multiscatter background populations are indicated with 1- and 2$\sigma$ contours. 
Without tagging scattered neutrons, PMS events and true single-scatter neutron interactions both produce NRs up to the maximum energy and contribute Migdal signals similarly. 
TPCMS interactions can produce large S2s, but their maximum S2 size is predicted to be approximately 50 electrons ($\sim$3.4 in the $\log_{10}(S2)$ space) as a result of the low neutron energy, allowing the predicted M-shell Migdal signals to dominate the parameter space with larger S2s. 
Neutron-induced gamma ray interactions in the TPC are also predicted to be subdominant. 
However, this simulation only considers near-beam neutrons ($<$\SI{5}{\degree}) and may have underestimated beam-related gamma ray backgrounds; an actual experiment must employ stringent gamma background controls to realize the anticipated Migdal sensitivity.  

To summarize, the NR Migdal effect can substantially improve the sensitivities of existing xenon experiments to sub-GeV dark matter candidates, which underscores the importance of its direct confirmation and characterization in liquid xenon. Drawing on our experimental efforts in searching for Migdal signals in liquid xenon, our modeling of xenon microphysics for signal generation, and recent simulation progress incorporating all these experimental effects, we predict that it is feasible to achieve a definitive Migdal effect measurement using neutron-induced xenon recoils. We are actively pursuing such an experiment with an expected M-shell Migdal signal-to-background ratio of $>$1, which would allow this effect to be quantified.

\begin{acknowledgments}
This work was performed under the auspices of the U.S. Department of Energy (DOE) by Lawrence Livermore National Laboratory (LLNL) under Contract DE-AC52-07NA27344. 
It is supported by the DOE Office of Science, Office of High Energy Physics under Work Proposal Numbers SCW1676 and SCW1930 awarded to LLNL. 
LLNL IM release number: LLNL-JRNL-2017344. 

The data that supports the findings of this study are available from the corresponding author upon reasonable request.
\end{acknowledgments}

\bibliographystyle{apsrev4-1}
\bibliography{biblio_shortened}

\begin{thebibliography}{34}%
\makeatletter
\providecommand \@ifxundefined [1]{%
 \@ifx{#1\undefined}
}%
\providecommand \@ifnum [1]{%
 \ifnum #1\expandafter \@firstoftwo
 \else \expandafter \@secondoftwo
 \fi
}%
\providecommand \@ifx [1]{%
 \ifx #1\expandafter \@firstoftwo
 \else \expandafter \@secondoftwo
 \fi
}%
\providecommand \natexlab [1]{#1}%
\providecommand \enquote  [1]{``#1''}%
\providecommand \bibnamefont  [1]{#1}%
\providecommand \bibfnamefont [1]{#1}%
\providecommand \citenamefont [1]{#1}%
\providecommand \href@noop [0]{\@secondoftwo}%
\providecommand \href [0]{\begingroup \@sanitize@url \@href}%
\providecommand \@href[1]{\@@startlink{#1}\@@href}%
\providecommand \@@href[1]{\endgroup#1\@@endlink}%
\providecommand \@sanitize@url [0]{\catcode `\\12\catcode `\$12\catcode
  `\&12\catcode `\#12\catcode `\^12\catcode `\_12\catcode `\%12\relax}%
\providecommand \@@startlink[1]{}%
\providecommand \@@endlink[0]{}%
\providecommand \url  [0]{\begingroup\@sanitize@url \@url }%
\providecommand \@url [1]{\endgroup\@href {#1}{\urlprefix }}%
\providecommand \urlprefix  [0]{URL }%
\providecommand \Eprint [0]{\href }%
\providecommand \doibase [0]{http://dx.doi.org/}%
\providecommand \selectlanguage [0]{\@gobble}%
\providecommand \bibinfo  [0]{\@secondoftwo}%
\providecommand \bibfield  [0]{\@secondoftwo}%
\providecommand \translation [1]{[#1]}%
\providecommand \BibitemOpen [0]{}%
\providecommand \bibitemStop [0]{}%
\providecommand \bibitemNoStop [0]{.\EOS\space}%
\providecommand \EOS [0]{\spacefactor3000\relax}%
\providecommand \BibitemShut  [1]{\csname bibitem#1\endcsname}%
\let\auto@bib@innerbib\@empty
\bibitem [{\citenamefont {Aalbers}\ \emph
  {et~al.}(2025{\natexlab{a}})\citenamefont {Aalbers} \emph
  {et~al.}}]{LZ2024_WS}%
  \BibitemOpen
  \bibfield  {author} {\bibinfo {author} {\bibfnamefont {J.}~\bibnamefont
  {Aalbers}} \emph {et~al.} (\bibinfo {collaboration} {LZ Collaboration}),\
  }\href {\doibase 10.1103/4dyc-z8zf} {\bibfield  {journal} {\bibinfo
  {journal} {Phys. Rev. Lett.}\ }\textbf {\bibinfo {volume} {135}},\ \bibinfo
  {pages} {011802} (\bibinfo {year} {2025}{\natexlab{a}})}\BibitemShut
  {NoStop}%
\bibitem [{\citenamefont {Aprile}\ \emph
  {et~al.}(2025{\natexlab{a}})\citenamefont {Aprile} \emph
  {et~al.}}]{XenonNT2025_NR}%
  \BibitemOpen
  \bibfield  {author} {\bibinfo {author} {\bibfnamefont {E.}~\bibnamefont
  {Aprile}} \emph {et~al.} (\bibinfo {collaboration} {XENON Collaboration}),\
  }\href {\doibase 10.1103/msw4-t342} {\bibfield  {journal} {\bibinfo
  {journal} {Phys. Rev. Lett.}\ }\textbf {\bibinfo {volume} {135}},\ \bibinfo
  {pages} {221003} (\bibinfo {year} {2025}{\natexlab{a}})}\BibitemShut
  {NoStop}%
\bibitem [{\citenamefont {Bo}\ \emph {et~al.}(2025)\citenamefont {Bo} \emph
  {et~al.}}]{PandaX2025_NRDM}%
  \BibitemOpen
  \bibfield  {author} {\bibinfo {author} {\bibfnamefont {Z.}~\bibnamefont {Bo}}
  \emph {et~al.} (\bibinfo {collaboration} {PandaX Collaboration}),\ }\href
  {\doibase 10.1103/PhysRevLett.134.011805} {\bibfield  {journal} {\bibinfo
  {journal} {Phys. Rev. Lett.}\ }\textbf {\bibinfo {volume} {134}},\ \bibinfo
  {pages} {011805} (\bibinfo {year} {2025})}\BibitemShut {NoStop}%
\bibitem [{\citenamefont {Aalbers}\ \emph
  {et~al.}(2025{\natexlab{b}})\citenamefont {Aalbers} \emph
  {et~al.}}]{XLZD2024_DesignBook}%
  \BibitemOpen
  \bibfield  {author} {\bibinfo {author} {\bibfnamefont {J.}~\bibnamefont
  {Aalbers}} \emph {et~al.} (\bibinfo {collaboration} {XLZD Collaboration}),\
  }\href {\doibase 10.1140/epjc/s10052-025-14810-w} {\bibfield  {journal}
  {\bibinfo  {journal} {The European Physical Journal C}\ }\textbf {\bibinfo
  {volume} {85}},\ \bibinfo {pages} {1} (\bibinfo {year}
  {2025}{\natexlab{b}})}\BibitemShut {NoStop}%
\bibitem [{\citenamefont {Essig}\ \emph
  {et~al.}(2012{\natexlab{a}})\citenamefont {Essig}, \citenamefont
  {Manalaysay}, \citenamefont {Mardon}, \citenamefont {Sorensen},\ and\
  \citenamefont {Volansky}}]{Essig2012_SubGeVXENON10}%
  \BibitemOpen
  \bibfield  {author} {\bibinfo {author} {\bibfnamefont {R.}~\bibnamefont
  {Essig}}, \bibinfo {author} {\bibfnamefont {A.}~\bibnamefont {Manalaysay}},
  \bibinfo {author} {\bibfnamefont {J.}~\bibnamefont {Mardon}}, \bibinfo
  {author} {\bibfnamefont {P.}~\bibnamefont {Sorensen}}, \ and\ \bibinfo
  {author} {\bibfnamefont {T.}~\bibnamefont {Volansky}},\ }\href {\doibase
  10.1103/PhysRevLett.109.021301} {\bibfield  {journal} {\bibinfo  {journal}
  {Phys. Rev. Lett.}\ }\textbf {\bibinfo {volume} {109}},\ \bibinfo {pages}
  {021301} (\bibinfo {year} {2012}{\natexlab{a}})}\BibitemShut {NoStop}%
\bibitem [{\citenamefont {Essig}\ \emph
  {et~al.}(2012{\natexlab{b}})\citenamefont {Essig}, \citenamefont {Mardon},\
  and\ \citenamefont {Volansky}}]{Essig2012_SubGeVDM}%
  \BibitemOpen
  \bibfield  {author} {\bibinfo {author} {\bibfnamefont {R.}~\bibnamefont
  {Essig}}, \bibinfo {author} {\bibfnamefont {J.}~\bibnamefont {Mardon}}, \
  and\ \bibinfo {author} {\bibfnamefont {T.}~\bibnamefont {Volansky}},\ }\href
  {\doibase 10.1103/PhysRevD.85.076007} {\bibfield  {journal} {\bibinfo
  {journal} {Phys. Rev. D}\ }\textbf {\bibinfo {volume} {85}},\ \bibinfo
  {pages} {076007} (\bibinfo {year} {2012}{\natexlab{b}})}\BibitemShut
  {NoStop}%
\bibitem [{\citenamefont {Akerib}\ \emph {et~al.}(2021)\citenamefont {Akerib}
  \emph {et~al.}}]{LUX_Migdal_2021}%
  \BibitemOpen
  \bibfield  {author} {\bibinfo {author} {\bibfnamefont {D.~S.}\ \bibnamefont
  {Akerib}} \emph {et~al.} (\bibinfo {collaboration} {LUX Collaboration}),\
  }\href {\doibase 10.1103/PhysRevD.104.012011} {\bibfield  {journal} {\bibinfo
   {journal} {Phys. Rev. D}\ }\textbf {\bibinfo {volume} {104}},\ \bibinfo
  {pages} {012011} (\bibinfo {year} {2021})}\BibitemShut {NoStop}%
\bibitem [{\citenamefont {Li}\ \emph {et~al.}(2023)\citenamefont {Li} \emph
  {et~al.}}]{pandax4t_s2o}%
  \BibitemOpen
  \bibfield  {author} {\bibinfo {author} {\bibfnamefont {S.}~\bibnamefont {Li}}
  \emph {et~al.} (\bibinfo {collaboration} {PandaX Collaboration}),\ }\href
  {http://dx.doi.org/10.1103/PhysRevLett.130.261001} {\bibfield  {journal}
  {\bibinfo  {journal} {Physical Review Letters}\ }\textbf {\bibinfo {volume}
  {130}} (\bibinfo {year} {2023})}\BibitemShut {NoStop}%
\bibitem [{\citenamefont {Lenardo}\ \emph {et~al.}(2019)\citenamefont
  {Lenardo}, \citenamefont {Xu}, \citenamefont {Pereverzev}, \citenamefont
  {Akindele}, \citenamefont {Naim}, \citenamefont {Kingston}, \citenamefont
  {Bernstein}, \citenamefont {Kazkaz}, \citenamefont {Tripathi}, \citenamefont
  {Awe}, \citenamefont {Li}, \citenamefont {Runge}, \citenamefont {Hedges},
  \citenamefont {An},\ and\ \citenamefont {Barbeau}}]{Lenardo2019_XeNR}%
  \BibitemOpen
  \bibfield  {author} {\bibinfo {author} {\bibfnamefont {B.~G.}\ \bibnamefont
  {Lenardo}}, \bibinfo {author} {\bibfnamefont {J.}~\bibnamefont {Xu}},
  \bibinfo {author} {\bibfnamefont {S.}~\bibnamefont {Pereverzev}}, \bibinfo
  {author} {\bibfnamefont {O.~A.}\ \bibnamefont {Akindele}}, \bibinfo {author}
  {\bibfnamefont {D.}~\bibnamefont {Naim}}, \bibinfo {author} {\bibfnamefont
  {J.}~\bibnamefont {Kingston}}, \bibinfo {author} {\bibfnamefont
  {A.}~\bibnamefont {Bernstein}}, \bibinfo {author} {\bibfnamefont
  {K.}~\bibnamefont {Kazkaz}}, \bibinfo {author} {\bibfnamefont
  {M.}~\bibnamefont {Tripathi}}, \bibinfo {author} {\bibfnamefont
  {C.}~\bibnamefont {Awe}}, \bibinfo {author} {\bibfnamefont {L.}~\bibnamefont
  {Li}}, \bibinfo {author} {\bibfnamefont {J.}~\bibnamefont {Runge}}, \bibinfo
  {author} {\bibfnamefont {S.}~\bibnamefont {Hedges}}, \bibinfo {author}
  {\bibfnamefont {P.}~\bibnamefont {An}}, \ and\ \bibinfo {author}
  {\bibfnamefont {P.~S.}\ \bibnamefont {Barbeau}},\ }\href {\doibase
  10.1103/PhysRevLett.123.231106} {\bibfield  {journal} {\bibinfo  {journal}
  {Phys. Rev. Lett.}\ }\textbf {\bibinfo {volume} {123}},\ \bibinfo {pages}
  {231106} (\bibinfo {year} {2019})}\BibitemShut {NoStop}%
\bibitem [{\citenamefont {Aprile}\ \emph
  {et~al.}(2025{\natexlab{b}})\citenamefont {Aprile} \emph
  {et~al.}}]{XENONnT2025_S2o}%
  \BibitemOpen
  \bibfield  {author} {\bibinfo {author} {\bibfnamefont {E.}~\bibnamefont
  {Aprile}} \emph {et~al.} (\bibinfo {collaboration} {XENON Collaboration}),\
  }\href {\doibase 10.1103/PhysRevLett.134.161004} {\bibfield  {journal}
  {\bibinfo  {journal} {Phys. Rev. Lett.}\ }\textbf {\bibinfo {volume} {134}},\
  \bibinfo {pages} {161004} (\bibinfo {year} {2025}{\natexlab{b}})}\BibitemShut
  {NoStop}%
\bibitem [{\citenamefont {Migdal}(1941)}]{Migdal_1941}%
  \BibitemOpen
  \bibfield  {author} {\bibinfo {author} {\bibfnamefont {A.~B.}\ \bibnamefont
  {Migdal}},\ }\href {\doibase
  www.itp.ac.ru/en/persons/migdal-arkady-beinusovich/} {\bibfield  {journal}
  {\bibinfo  {journal} {J. Phys. Acad. Sci. USSR}\ }\textbf {\bibinfo {volume}
  {4}},\ \bibinfo {pages} {449} (\bibinfo {year} {1941})}\BibitemShut {NoStop}%
\bibitem [{\citenamefont {Ibe}\ \emph {et~al.}(2018)\citenamefont {Ibe},
  \citenamefont {Nakano}, \citenamefont {Shoji},\ and\ \citenamefont
  {Suzuki}}]{Ibe2017_Migdal}%
  \BibitemOpen
  \bibfield  {author} {\bibinfo {author} {\bibfnamefont {M.}~\bibnamefont
  {Ibe}}, \bibinfo {author} {\bibfnamefont {W.}~\bibnamefont {Nakano}},
  \bibinfo {author} {\bibfnamefont {Y.}~\bibnamefont {Shoji}}, \ and\ \bibinfo
  {author} {\bibfnamefont {K.}~\bibnamefont {Suzuki}},\ }\href {\doibase
  10.1007/JHEP03(2018)194} {\bibfield  {journal} {\bibinfo  {journal} {Journal
  of High Energy Physics}\ }\textbf {\bibinfo {volume} {2018}},\ \bibinfo
  {pages} {194} (\bibinfo {year} {2018})}\BibitemShut {NoStop}%
\bibitem [{\citenamefont {Cox}\ \emph {et~al.}(2023)\citenamefont {Cox},
  \citenamefont {Dolan}, \citenamefont {McCabe},\ and\ \citenamefont
  {Quiney}}]{Cox2022_Migdal}%
  \BibitemOpen
  \bibfield  {author} {\bibinfo {author} {\bibfnamefont {P.}~\bibnamefont
  {Cox}}, \bibinfo {author} {\bibfnamefont {M.~J.}\ \bibnamefont {Dolan}},
  \bibinfo {author} {\bibfnamefont {C.}~\bibnamefont {McCabe}}, \ and\ \bibinfo
  {author} {\bibfnamefont {H.~M.}\ \bibnamefont {Quiney}},\ }\href {\doibase
  10.1103/PhysRevD.107.035032} {\bibfield  {journal} {\bibinfo  {journal}
  {Phys. Rev. D}\ }\textbf {\bibinfo {volume} {107}},\ \bibinfo {pages}
  {035032} (\bibinfo {year} {2023})}\BibitemShut {NoStop}%
\bibitem [{\citenamefont {Essig}\ \emph {et~al.}(2020)\citenamefont {Essig},
  \citenamefont {Pradler}, \citenamefont {Sholapurkar},\ and\ \citenamefont
  {Yu}}]{Essig2020:Migdal}%
  \BibitemOpen
  \bibfield  {author} {\bibinfo {author} {\bibfnamefont {R.}~\bibnamefont
  {Essig}}, \bibinfo {author} {\bibfnamefont {J.}~\bibnamefont {Pradler}},
  \bibinfo {author} {\bibfnamefont {M.}~\bibnamefont {Sholapurkar}}, \ and\
  \bibinfo {author} {\bibfnamefont {T.-T.}\ \bibnamefont {Yu}},\ }\href
  {\doibase 10.1103/PhysRevLett.124.021801} {\bibfield  {journal} {\bibinfo
  {journal} {Phys. Rev. Lett.}\ }\textbf {\bibinfo {volume} {124}},\ \bibinfo
  {pages} {021801} (\bibinfo {year} {2020})}\BibitemShut {NoStop}%
\bibitem [{\citenamefont {Liu}\ \emph {et~al.}(2020)\citenamefont {Liu},
  \citenamefont {Wu}, \citenamefont {Chi},\ and\ \citenamefont
  {Chen}}]{Liu2020_Migdal}%
  \BibitemOpen
  \bibfield  {author} {\bibinfo {author} {\bibfnamefont {C.-P.}\ \bibnamefont
  {Liu}}, \bibinfo {author} {\bibfnamefont {C.-P.}\ \bibnamefont {Wu}},
  \bibinfo {author} {\bibfnamefont {H.-C.}\ \bibnamefont {Chi}}, \ and\
  \bibinfo {author} {\bibfnamefont {J.-W.}\ \bibnamefont {Chen}},\ }\href
  {\doibase 10.1103/PhysRevD.102.121303} {\bibfield  {journal} {\bibinfo
  {journal} {Phys. Rev. D}\ }\textbf {\bibinfo {volume} {102}},\ \bibinfo
  {pages} {121303} (\bibinfo {year} {2020})}\BibitemShut {NoStop}%
\bibitem [{\citenamefont {Baxter}\ \emph {et~al.}(2020)\citenamefont {Baxter},
  \citenamefont {Kahn},\ and\ \citenamefont {Krnjaic}}]{Baxter2020:Migdal}%
  \BibitemOpen
  \bibfield  {author} {\bibinfo {author} {\bibfnamefont {D.}~\bibnamefont
  {Baxter}}, \bibinfo {author} {\bibfnamefont {Y.}~\bibnamefont {Kahn}}, \ and\
  \bibinfo {author} {\bibfnamefont {G.}~\bibnamefont {Krnjaic}},\ }\href
  {\doibase 10.1103/PhysRevD.101.076014} {\bibfield  {journal} {\bibinfo
  {journal} {Phys. Rev. D}\ }\textbf {\bibinfo {volume} {101}},\ \bibinfo
  {pages} {076014} (\bibinfo {year} {2020})}\BibitemShut {NoStop}%
\bibitem [{\citenamefont {Boehm}\ and\ \citenamefont
  {Wu}(1954)}]{Boehm1954_BetaMigdal}%
  \BibitemOpen
  \bibfield  {author} {\bibinfo {author} {\bibfnamefont {F.}~\bibnamefont
  {Boehm}}\ and\ \bibinfo {author} {\bibfnamefont {C.~S.}\ \bibnamefont {Wu}},\
  }\href {\doibase 10.1103/PhysRev.93.518} {\bibfield  {journal} {\bibinfo
  {journal} {Phys. Rev.}\ }\textbf {\bibinfo {volume} {93}},\ \bibinfo {pages}
  {518} (\bibinfo {year} {1954})}\BibitemShut {NoStop}%
\bibitem [{\citenamefont {Rapaport}\ \emph {et~al.}(1975)\citenamefont
  {Rapaport}, \citenamefont {Asaro},\ and\ \citenamefont
  {Perlman}}]{Rapaport1975_AlphaMigdal}%
  \BibitemOpen
  \bibfield  {author} {\bibinfo {author} {\bibfnamefont {M.~S.}\ \bibnamefont
  {Rapaport}}, \bibinfo {author} {\bibfnamefont {F.}~\bibnamefont {Asaro}}, \
  and\ \bibinfo {author} {\bibfnamefont {I.}~\bibnamefont {Perlman}},\ }\href
  {\doibase 10.1103/PhysRevC.11.1740} {\bibfield  {journal} {\bibinfo
  {journal} {Phys. Rev. C}\ }\textbf {\bibinfo {volume} {11}},\ \bibinfo
  {pages} {1740} (\bibinfo {year} {1975})}\BibitemShut {NoStop}%
\bibitem [{\citenamefont {Yi}\ \emph {et~al.}(2026)\citenamefont {Yi} \emph
  {et~al.}}]{Yi2026_Migdal}%
  \BibitemOpen
  \bibfield  {author} {\bibinfo {author} {\bibfnamefont {D.}~\bibnamefont {Yi}}
  \emph {et~al.},\ }\href {\doibase 10.1038/s41586-025-09918-8} {\bibfield
  {journal} {\bibinfo  {journal} {Nature}\ }\textbf {\bibinfo {volume} {649}},\
  \bibinfo {pages} {580} (\bibinfo {year} {2026})}\BibitemShut {NoStop}%
\bibitem [{\citenamefont {Akerib}\ \emph {et~al.}(2019)\citenamefont {Akerib}
  \emph {et~al.}}]{LUX2019:Migdal}%
  \BibitemOpen
  \bibfield  {author} {\bibinfo {author} {\bibfnamefont {D.~S.}\ \bibnamefont
  {Akerib}} \emph {et~al.} (\bibinfo {collaboration} {LUX Collaboration}),\
  }\href {\doibase 10.1103/PhysRevLett.122.131301} {\bibfield  {journal}
  {\bibinfo  {journal} {Phys. Rev. Lett.}\ }\textbf {\bibinfo {volume} {122}},\
  \bibinfo {pages} {131301} (\bibinfo {year} {2019})}\BibitemShut {NoStop}%
\bibitem [{\citenamefont {Aprile}\ \emph {et~al.}(2019)\citenamefont {Aprile}
  \emph {et~al.}}]{XENON1T2019:Migdal}%
  \BibitemOpen
  \bibfield  {author} {\bibinfo {author} {\bibfnamefont {E.}~\bibnamefont
  {Aprile}} \emph {et~al.} (\bibinfo {collaboration} {XENON Collaboration}),\
  }\href {\doibase 10.1103/PhysRevLett.123.241803} {\bibfield  {journal}
  {\bibinfo  {journal} {Phys. Rev. Lett.}\ }\textbf {\bibinfo {volume} {123}},\
  \bibinfo {pages} {241803} (\bibinfo {year} {2019})}\BibitemShut {NoStop}%
\bibitem [{\citenamefont {Huang}\ \emph {et~al.}(2023)\citenamefont {Huang}
  \emph {et~al.}}]{PandaX2023_Migdal}%
  \BibitemOpen
  \bibfield  {author} {\bibinfo {author} {\bibfnamefont {D.}~\bibnamefont
  {Huang}} \emph {et~al.} (\bibinfo {collaboration} {PandaX Collaboration}),\
  }\href {\doibase 10.1103/PhysRevLett.131.191002} {\bibfield  {journal}
  {\bibinfo  {journal} {Phys. Rev. Lett.}\ }\textbf {\bibinfo {volume} {131}},\
  \bibinfo {pages} {191002} (\bibinfo {year} {2023})}\BibitemShut {NoStop}%
\bibitem [{\citenamefont {Araújo}\ \emph {et~al.}(2023)\citenamefont {Araújo}
  \emph {et~al.}}]{UKMIGDAL2022}%
  \BibitemOpen
  \bibfield  {author} {\bibinfo {author} {\bibfnamefont {H.}~\bibnamefont
  {Araújo}} \emph {et~al.},\ }\href {\doibase
  https://doi.org/10.1016/j.astropartphys.2023.102853} {\bibfield  {journal}
  {\bibinfo  {journal} {Astroparticle Physics}\ }\textbf {\bibinfo {volume}
  {151}},\ \bibinfo {pages} {102853} (\bibinfo {year} {2023})}\BibitemShut
  {NoStop}%
\bibitem [{\citenamefont {Xu}\ \emph {et~al.}(2024)\citenamefont {Xu},
  \citenamefont {Adams}, \citenamefont {Lenardo}, \citenamefont {Pershing},
  \citenamefont {Mannino}, \citenamefont {Bernard}, \citenamefont {Kingston},
  \citenamefont {Mizrachi}, \citenamefont {Lin}, \citenamefont {Essig} \emph
  {et~al.}}]{Xu2024_Migdal}%
  \BibitemOpen
  \bibfield  {author} {\bibinfo {author} {\bibfnamefont {J.}~\bibnamefont
  {Xu}}, \bibinfo {author} {\bibfnamefont {D.}~\bibnamefont {Adams}}, \bibinfo
  {author} {\bibfnamefont {B.}~\bibnamefont {Lenardo}}, \bibinfo {author}
  {\bibfnamefont {T.}~\bibnamefont {Pershing}}, \bibinfo {author}
  {\bibfnamefont {R.}~\bibnamefont {Mannino}}, \bibinfo {author} {\bibfnamefont
  {E.}~\bibnamefont {Bernard}}, \bibinfo {author} {\bibfnamefont
  {J.}~\bibnamefont {Kingston}}, \bibinfo {author} {\bibfnamefont
  {E.}~\bibnamefont {Mizrachi}}, \bibinfo {author} {\bibfnamefont
  {J.}~\bibnamefont {Lin}}, \bibinfo {author} {\bibfnamefont {R.}~\bibnamefont
  {Essig}},  \emph {et~al.},\ }\href {\doibase
  https://doi.org/10.1103/PhysRevD.109.L051101} {\bibfield  {journal} {\bibinfo
   {journal} {Physical Review D}\ }\textbf {\bibinfo {volume} {109}},\ \bibinfo
  {pages} {L051101} (\bibinfo {year} {2024})}\BibitemShut {NoStop}%
\bibitem [{\citenamefont {Bang}(2024)}]{LZ_Bang_phd}%
  \BibitemOpen
  \bibfield  {author} {\bibinfo {author} {\bibfnamefont {J.}~\bibnamefont
  {Bang}},\ }\emph {\bibinfo {title} {{Implementation of a Deuterium-Deuterium
  Neutron Generator and its Application for Nuclear Recoil Calibration and the
  Search for the Migdal Effect in the LUX-ZEPLIN Experiment}}},\ \href
  {https://repository.library.brown.edu/studio/item/bdr:37s8v2v2/} {Ph.D.
  thesis},\ \bibinfo  {school} {Brown University} (\bibinfo {year}
  {2024})\BibitemShut {NoStop}%
\bibitem [{\citenamefont {Vaitkus}(2025)}]{LZ_Vaitkus_phd}%
  \BibitemOpen
  \bibfield  {author} {\bibinfo {author} {\bibfnamefont {A.}~\bibnamefont
  {Vaitkus}},\ }\emph {\bibinfo {title} {{Characterization of the R11410-22
  Photomultiplier Tube and the Search for a Neutron-Induced Migdal Effect in
  the LUX-ZEPLIN Experiment}}},\ \href
  {https://repository.library.brown.edu/studio/item/bdr:7jaat7fh/} {Ph.D.
  thesis},\ \bibinfo  {school} {Brown University} (\bibinfo {year}
  {2025})\BibitemShut {NoStop}%
\bibitem [{\citenamefont {Nakamura}\ \emph {et~al.}(2020)\citenamefont
  {Nakamura}, \citenamefont {Miuchi}, \citenamefont {Kazama}, \citenamefont
  {Shoji}, \citenamefont {Ibe},\ and\ \citenamefont
  {Nakano}}]{Nakamura2020_Migdal}%
  \BibitemOpen
  \bibfield  {author} {\bibinfo {author} {\bibfnamefont {K.~D.}\ \bibnamefont
  {Nakamura}}, \bibinfo {author} {\bibfnamefont {K.}~\bibnamefont {Miuchi}},
  \bibinfo {author} {\bibfnamefont {S.}~\bibnamefont {Kazama}}, \bibinfo
  {author} {\bibfnamefont {Y.}~\bibnamefont {Shoji}}, \bibinfo {author}
  {\bibfnamefont {M.}~\bibnamefont {Ibe}}, \ and\ \bibinfo {author}
  {\bibfnamefont {W.}~\bibnamefont {Nakano}},\ }\href {\doibase
  10.1093/ptep/ptaa162} {\bibfield  {journal} {\bibinfo  {journal} {Progress of
  Theoretical and Experimental Physics}\ }\textbf {\bibinfo {volume} {2021}},\
  \bibinfo {pages} {013C01} (\bibinfo {year} {2020})}\BibitemShut {NoStop}%
\bibitem [{\citenamefont {Xu}\ \emph {et~al.}(2023)\citenamefont {Xu},
  \citenamefont {Barbeau},\ and\ \citenamefont {Hong}}]{Xu2023_NRCalibration}%
  \BibitemOpen
  \bibfield  {author} {\bibinfo {author} {\bibfnamefont {J.}~\bibnamefont
  {Xu}}, \bibinfo {author} {\bibfnamefont {P.}~\bibnamefont {Barbeau}}, \ and\
  \bibinfo {author} {\bibfnamefont {Z.}~\bibnamefont {Hong}},\ }\href {\doibase
  https://doi.org/10.1146/annurev-nucl-111722-025122} {\bibfield  {journal}
  {\bibinfo  {journal} {Annual Review of Nuclear and Particle Science}\
  }\textbf {\bibinfo {volume} {73}},\ \bibinfo {pages} {95} (\bibinfo {year}
  {2023})}\BibitemShut {NoStop}%
\bibitem [{\citenamefont {Akerib}\ \emph {et~al.}(2025)\citenamefont {Akerib}
  \emph {et~al.}}]{LUX2025_NR}%
  \BibitemOpen
  \bibfield  {author} {\bibinfo {author} {\bibfnamefont {D.~S.}\ \bibnamefont
  {Akerib}} \emph {et~al.} (\bibinfo {collaboration} {LUX Collaboration}),\
  }\href {\doibase 10.1103/PhysRevLett.134.061002} {\bibfield  {journal}
  {\bibinfo  {journal} {Phys. Rev. Lett.}\ }\textbf {\bibinfo {volume} {134}},\
  \bibinfo {pages} {061002} (\bibinfo {year} {2025})}\BibitemShut {NoStop}%
\bibitem [{\citenamefont {Albakry}\ \emph {et~al.}(2023)\citenamefont {Albakry}
  \emph {et~al.}}]{CDMS2023_SiNR}%
  \BibitemOpen
  \bibfield  {author} {\bibinfo {author} {\bibfnamefont {M.~F.}\ \bibnamefont
  {Albakry}} \emph {et~al.} (\bibinfo {collaboration} {SuperCDMS
  Collaboration}),\ }\href {\doibase 10.1103/PhysRevLett.131.091801} {\bibfield
   {journal} {\bibinfo  {journal} {Phys. Rev. Lett.}\ }\textbf {\bibinfo
  {volume} {131}},\ \bibinfo {pages} {091801} (\bibinfo {year}
  {2023})}\BibitemShut {NoStop}%
\bibitem [{\citenamefont {Temples}\ \emph {et~al.}(2021)\citenamefont
  {Temples}, \citenamefont {McLaughlin}, \citenamefont {Bargemann},
  \citenamefont {Baxter}, \citenamefont {Cottle}, \citenamefont {Dahl},
  \citenamefont {Lippincott}, \citenamefont {Monte},\ and\ \citenamefont
  {Phelan}}]{Temples2021_127Xe}%
  \BibitemOpen
  \bibfield  {author} {\bibinfo {author} {\bibfnamefont {D.~J.}\ \bibnamefont
  {Temples}}, \bibinfo {author} {\bibfnamefont {J.}~\bibnamefont {McLaughlin}},
  \bibinfo {author} {\bibfnamefont {J.}~\bibnamefont {Bargemann}}, \bibinfo
  {author} {\bibfnamefont {D.}~\bibnamefont {Baxter}}, \bibinfo {author}
  {\bibfnamefont {A.}~\bibnamefont {Cottle}}, \bibinfo {author} {\bibfnamefont
  {C.~E.}\ \bibnamefont {Dahl}}, \bibinfo {author} {\bibfnamefont {W.~H.}\
  \bibnamefont {Lippincott}}, \bibinfo {author} {\bibfnamefont
  {A.}~\bibnamefont {Monte}}, \ and\ \bibinfo {author} {\bibfnamefont
  {J.}~\bibnamefont {Phelan}},\ }\href {\doibase 10.1103/PhysRevD.104.112001}
  {\bibfield  {journal} {\bibinfo  {journal} {Phys. Rev. D}\ }\textbf {\bibinfo
  {volume} {104}},\ \bibinfo {pages} {112001} (\bibinfo {year}
  {2021})}\BibitemShut {NoStop}%
\bibitem [{\citenamefont {Aalbers}\ \emph
  {et~al.}(2025{\natexlab{c}})\citenamefont {Aalbers} \emph
  {et~al.}}]{LZ2025_EC}%
  \BibitemOpen
  \bibfield  {author} {\bibinfo {author} {\bibfnamefont {J.}~\bibnamefont
  {Aalbers}} \emph {et~al.} (\bibinfo {collaboration} {LZ Collaboration}),\
  }\href {\doibase 10.1103/447w-94h3} {\bibfield  {journal} {\bibinfo
  {journal} {Phys. Rev. D}\ }\textbf {\bibinfo {volume} {112}},\ \bibinfo
  {pages} {012024} (\bibinfo {year} {2025}{\natexlab{c}})}\BibitemShut
  {NoStop}%
\bibitem [{\citenamefont {Xu}\ \emph {et~al.}(2025)\citenamefont {Xu} \emph
  {et~al.}}]{Xu2025_Recombination}%
  \BibitemOpen
  \bibfield  {author} {\bibinfo {author} {\bibfnamefont {J.}~\bibnamefont {Xu}}
  \emph {et~al.},\ }\href {\doibase 10.1103/8bdr-6njs} {\bibfield  {journal}
  {\bibinfo  {journal} {Phys. Rev. D}\ }\textbf {\bibinfo {volume} {112}},\
  \bibinfo {pages} {012014} (\bibinfo {year} {2025})}\BibitemShut {NoStop}%
\bibitem [{\citenamefont {Szydagis}\ \emph {et~al.}(2025)\citenamefont
  {Szydagis} \emph {et~al.}}]{Szydagis2025_NEST}%
  \BibitemOpen
  \bibfield  {author} {\bibinfo {author} {\bibfnamefont {M.}~\bibnamefont
  {Szydagis}} \emph {et~al.},\ }\href
  {https://www.frontiersin.org/journals/detector-science-and-technology/articles/10.3389/fdest.2024.1480975}
  {\bibfield  {journal} {\bibinfo  {journal} {Frontiers in Detector Science and
  Technology}\ }\textbf {\bibinfo {volume} {Volume 2 - 2024}} (\bibinfo {year}
  {2025})}\BibitemShut {NoStop}%
\end{thebibliography}%

\end{document}